\documentclass{article}
\usepackage{graphicx}
\usepackage{amsmath}

\begin{document}

\title{Trace anomaly and black holes evaporation }
\author{F. Canfora and G. Vilasi\thanks{%
Corresponding author: Gaetano Vilasi, Universit\`{a} di Salerno,
Dipartimento di Fisica \textit{E.R.Caianiello}, Via S.Allende, 84081
Baronissi (Salerno) Italy; e-mail: vilasi@sa.infn.it, Phone: +39-089-965317,
Fax: +39-089-965275.} \\
\emph{Istituto Nazionale di Fisica Nucleare, GC di Salerno, Italy.}\textit{\
}\\
\emph{Dipartimento di Fisica ''E.R.Caianiello'', Universit\`{a} di Salerno.}%
\\
}
\maketitle

\begin{abstract}
A model is proposed to describe a transition from a Schwarzschild black hole
of mass $M_{0}$ to a Schwarzschild black hole of mass $M_{1}$ $\leq M_{0}$.
The basic equations are derived from the non-vacuum Einstein field equations
taking a source representing a null scalar field with a nonvanishing trace
anomaly. It is shown that the nonvanishing trace anomaly of the scalar field
prevents a complete evaporation.

\textit{PACS numbers: 04.70.Dy, 04.20Gz}
\end{abstract}

In the Hawking seminal paper in 1974 \cite{Ha74} it was shown that a
Schwarzschild black-hole, the most\textit{\ attractive} object of
theoretical physics, does emit thermal radiation. This discovery provided
deeper physical basis to the Bekenstein entropy \cite{Be73} and to the
black-hole thermodynamics.

However, this new understanding of the black hole physics raised some highly
nontrivial questions about the foundations of Quantum Field Theory (QFT). In
particular, the black-hole evaporation led to an apparent breakdown of
predictability \cite{Ha76} because of the thermal character of the emitted
flux. Eventually, if the evaporation is complete, a loss of information and
of unitarity occur since, in this case, a pure state evolves in a density
matrix. Among all these proposals to overcome these difficulties, probably
the three most popular are:

\begin{itemize}
\item[(i)]  QFT has to be generalized to allow nonunitary processes \cite
{Ha76} since it seems that a complete evaporation cannot be ruled out.
However, up to now, this idea has been found to be incompatible with
locality or conservation of energy \cite{El84}, \cite{Ba84};

\item[(ii)]  a Planckian remnant, which could store the lost information, is
left \cite{Ah87}. Moreover, as showed by Bekenstein \cite{Be93}, the actual
information loss due to the Hawking radiation is lower then that of a pure
blackbody. Thus, a part of lost information could be detained by the remnant
itself and the remaining part could be encoded in the Hawking radiation
through the Bekenstein mechanism;

\item[(iii)]  a quantum theory of gravity, such as \textit{superstring}
\textit{theory} \cite{Da02} or \textit{loop} \textit{quantum} \textit{gravity%
} \cite{Ti01}, will disclose the information hidden in the detailed
structure of the emitted quanta \cite{Pa80}.
\end{itemize}

Here, a model is proposed which shows that a complete evaporation does not
occur if one takes into account the trace anomaly of the null quantum fields
that carry away energy from the black hole to infinity. Furthermore, the
approach sheds light on the relation between the conservation of energy and
the black hole evaporation.

\textit{The model \ \ }It is believed that a full understanding of the black
hole evaporation requires a full quantum theory of gravity such as the
\textit{superstring} \textit{theory} or the \textit{loop} \textit{quantum}
\textit{gravity}. In recent years, even more ''conventional'' approaches to
the quantum gravity, such as the \textit{effective} \textit{action} and the
\textit{renormalization} \textit{group} \textit{method} \cite{Bo00}, gave
new hints for approaching this problem. However, until the final stages,
this is unnecessary because a black hole evaporation cannot be considered a
strong gravitational field phenomenon since there is a slow rate of change
of the black hole mass with time for almost all the process \cite{Ba81}.
Then, it must be possible to describe how the metric changes, owing to the
evaporation, by the standard Einstein equations with a suitable matter
source. Furthermore, a very interesting application of \textit{loop} \textit{%
quantum} \textit{gravity}, shows that the discrete eigenvalues of the
geometric operators converge very fast, actually in a few Plank lengths, to
their semiclassical approximations \cite{Bo01l} \cite{Bo01}.

For this reason we can assume that, even if near the horizon there could be
quantum gravity effects, far enough (with respect to the Planck length) from
the horizon a semiclassical description should give the right results.
Therefore, the semiclassical framework is well suited to study the blackhole
evaporation, since, to study the evaporation, one has to look at the energy
radiated at infinity. Moreover, it is desirable for the self-consistency of
QFT to solve the problems raised by the blackhole evaporation by only using
general relativity and QFT itself.

For these reasons, some authors claimed that the issues raised by Hawking
radiation could be resolved by including in the models the back-reaction
effects in the framework of the \textit{semiclassical program} \cite{Ba81},
\cite{Pa82}, \cite{Yo85}, \cite{Ah87}, \cite{Pa94}, \cite{Ma95}, \cite{Ma00}%
, \cite{Ca01}. The standard \textit{semiclassical program} for studying the
back-reaction in the black-hole evaporation consists, first, in evaluating
an effective vacuum stress-energy tensor of a null quantum field on the
black hole space-time of interest \cite{Pa82} and, then, using this tensor
as a source for the Einstein equations \cite{Yo85}.

This kind of approach has the advantage that results are easy to interpret
in terms of the unperturbed quantity, thus making transparent the influence
of the source on the back-ground geometry \cite{Yo85}.

However, the calculation of the vacuum stress-energy tensor of quantum
fields on a curved space-time is not easy. The approximated form of the
stress-energy tensor is computed in a four-dimensional geometry only in
static situations \cite{Pa82}, while, in non static models, it is usually
computed in effective two-dimensional geometries \cite{Pa94, GKV02}.
Therefore, in such effective models, it is not possible to take into account
the nontrivial effects of the four-dimensional trace anomaly\footnote{%
Some attempts to take into account the four-dimensional trace anomaly in the
framework of \textit{dilatonic gravity} can be found in \cite{No00}.}.

Thus, an approach is developed in which the form of the stress-energy tensor
is kept general but the trace anomaly is fully taken into account. It will
be shown, \textit{via} the exact Einstein equations, that this has important
physical consequences.

\textit{The Einstein tensor \ \ }A transition from a Schwarzschild blackhole
of initial mass $M$ to a Schwarzschild blackhole of final mass $\overline{M}$
will be described outside the collapsed matter by the means of a well suited
source that carries away energy from the blackhole to infinity.\ The metric,
valid only for $r>R^{\ast }>0$, $R^{\ast }$ denoting the initial radius of
the horizon, will be written as in \cite{Ba81}:

\begin{equation}
ds^{2}=\left( 1-\frac{m(r,v)}{r}\right)
e^{2j(r,v)}dv^{2}-2e^{j(r,v)}drdv-r^{2}d\Omega \text{.}  \label{m1}
\end{equation}
For $m=const$ and $j=const$, metric (\ref{m1}) reduces to the Schwarzschild
metric in Eddington-Finkelstein coordinates.

Thus, in order to describe a transition from a Schwarzschild black hole of
mass $M$ to a Schwarzschild black hole of mass $\overline{M}$, we will
search for a solution of the Einstein equations satisfying the following
boundary conditions:
\begin{equation*}
m(r,0)=M>0;\,m(r,T)=\overline{M}\geq 0;\,j(r,0)=k_{1};\,j(r,T)=k_{2},
\end{equation*}
where $k_{1}$ and $k_{2}$\ are constants and at $v=T$ \ the blackhole
reaches its final state.

The above metric has the following scalar curvature
\begin{align*}
R& =\left\{ -\exp (-j)2r^{2}\partial _{rv}^{2}j+3r\partial _{r}j\partial
_{r}m+m\partial _{r}j\right. \\
& +2\partial _{r}m-4r\partial _{r}j-2r^{2}\left( 1-\frac{m}{r}\right)
\partial _{r}^{2}j \\
& +\left. -2r^{2}\left( 1-\frac{m}{r}\right) \left( \partial _{r}j\right)
^{2}+r\partial _{r}^{2}m\right\} \frac{1}{r^{2}}
\end{align*}
and the nonvanishing independent components of the Einstein tensor are
\begin{align*}
G_{11}& =\frac{\exp \left( 2j\right) }{r^{3}}\left( m\partial _{r}m-r\exp
(-j)\partial _{v}m-r\partial _{r}m\right) , \\
G_{12}& =\frac{\exp (j)\partial _{r}m}{r^{2}},\text{ \ \ \ \ \ }G_{22}=-2%
\frac{\partial _{r}j}{r}, \\
G_{33}& =\frac{1}{2}\left\{ -\left( 2r+m\right) \partial _{r}j-\exp
(-j)2r^{2}\partial _{rv}^{2}j\right. \\
& +3r\partial _{r}j\partial _{r}m-2r^{2}\left( 1-\frac{m}{r}\right) \partial
_{r}^{2}j \\
& \left. -2r^{2}\left( 1-\frac{m}{r}\right) \left( \partial _{r}j\right)
^{2}+r\partial _{r}^{2}m\right\} ,\text{ } \\
G_{44}& =\sin ^{2}\theta G_{33}\text{.}
\end{align*}

\textit{The source \ \ }The commonly accepted picture describing the
evaporation is that the Schwarzschild black-hole emits black-body radiation
at the Hawking temperature. Since the emission of massive particles by the
blackhole is highly suppressed, the natural choice is to take as source a
null\ quantum fluid, \textit{i.e.} a null fluid violating all the energy
conditions:
\begin{equation*}
T_{\mu \nu }=g_{\mu \nu }V\left( v,r\right) +\rho \left( v,r\right) U_{\mu
}U_{\nu }\text{.}
\end{equation*}
Thus, $T_{\mu \nu }$ has to be interpreted as a suitable average over
quantum fields that share to Hawking radiation and take away energy from the
collapsing matter to infinity, while $V$ corresponds to the trace anomaly of
quantum fields. Since no constraints will be imposed on $\rho $ and $V$, the
tensor $T_{\mu \nu }$ can describe a classical, a semiclassical or a purely
quantum source, so that neither the \textit{weak} nor the \textit{null}
energy conditions need to be satisfied. When $\rho \geq 0$, $T_{\mu \nu }$
can be thought as the energy-momentum tensor of a null scalar field with
trace anomaly and, in this case, the \textit{weak} energy condition can
still be violated but the \textit{null} cannot. At a first glance it seems
that we would have to specify the explicit form of $\rho $ and $V$, but the
Einstein field equations will give a closed system of equations for $m(v,r)$
and $j(v,r)$ which does not depend on the detailed form of $T_{\mu \nu }$.
Thus, the gravitational part ''decouples'' from the matter source. Thanks to
this fact, it is possible to keep general the form of $T_{\mu \nu }$.

\textit{The Einstein equations \ \ }taking into account that $g_{\mu \nu
}U^{\mu }U^{\nu }=0$, \textit{\ }the Einstein equations $G_{\mu \nu }=\kappa
T_{\mu \nu }$ give:
\begin{align}
\kappa ^{-1}G_{11}& =V\left( 1-\frac{m}{r}\right) \exp \left( 2j\right)
+\rho \left( U_{v}\right) ^{2},  \notag \\
\kappa ^{-1}G_{12}& =-\exp \left( j\right) V+\rho U_{r}U_{v},\text{ \ }%
\kappa ^{-1}G_{22}=\rho \left( U_{r}\right) ^{2},  \label{stop} \\
\kappa ^{-1}G_{33}& =g_{33}V,\text{ \ }-R=4\kappa V.  \notag
\end{align}

The last two equations give for the scalar curvature
\begin{equation*}
R=\frac{4}{r^{2}}\left[ \left( m-r\right) \partial _{r}j+\partial _{r}m%
\right] .
\end{equation*}
Moreover, from
\begin{equation*}
G_{12}^{2}-G_{11}G_{22}=\kappa ^{2}\left( T_{12}^{2}-T_{11}T_{22}\right)
=\kappa ^{2}\left( g_{12}\right) ^{2}V^{2}
\end{equation*}
it follows that
\begin{equation*}
\left( m-r\right) ^{2}\left( \partial _{r}j\right) ^{2}=-2r\exp \left(
-j\right) \partial _{r}j\partial _{t}m\text{.}
\end{equation*}
Thus, by assuming\footnote{%
It can be easily seen that $\partial _{r}j=0$ gives only static solutions,
or solutions with vanishing trace anomaly.} $\partial _{r}j\neq 0$, the
equations for $m$ and $j$ reduce to
\begin{equation}
\frac{\exp \left( j\right) \partial _{r}j}{2}=\partial _{v}\left( \frac{r}{%
m-r}\right) ,  \label{F1}
\end{equation}
\begin{eqnarray}
3m\partial _{r}j &=&-\exp \left( -j\right) 2r^{2}\partial _{rv}^{2}j+\left(
3r\partial _{r}j-2\right) \partial _{r}m  \notag \\
&&-r\partial _{r}^{2}m-2r^{2}\left( 1-\frac{m}{r}\right) \left( \partial
_{r}^{2}j+\left( \partial _{r}j\right) ^{2}\right) ,  \label{F2}
\end{eqnarray}
where it is worth to note that only Eq.(\ref{F1}) depends on the assumption $%
\partial _{r}j\neq 0$. Thus, the problem reduces to solve Eqs. (\ref{F1})
and (\ref{F2}) with the boundary conditions
\begin{align}
m,j& \in C^{2}\left( \left[ 0,T\right] \times \left[ R^{\ast },\infty \right[
\right) ,\text{ \ \ }  \label{derivabilità} \\
\left| m\right| ,\left| j\right| & \leq K_{1}<+\infty ,\text{ \ \ \ }\left|
\partial _{v}j\right| ,\left| \partial _{r}j\right| \underset{r\rightarrow
+\infty }{\leq }K_{2}r^{-2}  \label{ADM} \\
m(0,r)& =M>0;\,j(0,r)=k_{1};\,m(T,r)=\overline{M}\geq 0;j(T,r)=k_{2},\text{%
\quad }
\end{align}
where the upper bounds (\ref{ADM}) are needed to have an asymptotically flat
metric. It is important to stress that Eqs (\ref{F1}) and (\ref{F2}), not
depending on the explicit form of $\rho $ and $V$, are decoupled from the
matter source but, however, are a direct consequence of the non vanishing
trace anomaly.

\textit{Estimate of the final mass \ \ }Now, it will be proven that the
trace anomaly prevents a complete evaporation. Indeed, by using the boundary
conditions $j(0,r)=0$ and $m(0,r)=M$, from Eq. (\ref{F1}) it follows that
the function $m$ can be expressed in closed form in terms of $j$ in the
following way:
\begin{equation}
m(v,r)=r-\frac{r}{\frac{r}{r-M}-\frac{1}{2}\partial _{r}\int_{0}^{v}\exp %
\left[ j(\tau ,r)\right] d\tau },  \label{rapu}
\end{equation}
so that the final mass $\overline{M}$ may be expressed as
\begin{equation}
\overline{M}=r-\frac{r}{\frac{r}{r-M}-\frac{1}{2}\partial
_{r}\int_{0}^{T}\exp \left[ j(\tau ,r)\right] d\tau }.  \label{FM}
\end{equation}

It is worth to note that $m(v,r)$, and then $\overline{M}$,\ cannot be
negative. In fact, by using Eqs (\ref{ADM}) and (\ref{FM}) we have:
\begin{equation*}
m\underset{r\rightarrow +\infty }{\sim }r-\frac{r}{1+\frac{M}{r}-v\frac{%
K_{2}\exp \left[ K_{1}\right] }{2r^{2}}}\geq 0
\end{equation*}
being $M>0$. Furthermore, by using Eqs. (\ref{stop}) it is trivial to show
that:
\begin{align*}
\rho & \geq 0\Rightarrow \partial _{r}j\leq 0\Rightarrow \overline{M}\geq M
\\
\rho & <0\Rightarrow \partial _{r}j>0\Rightarrow \overline{M}<M\text{.}
\end{align*}
Then, when $\rho \geq 0$, so that $T_{\mu \nu }$ describes a null scalar
field violating the \textit{weak} energy condition but not the \textit{null}
one, the evaporation is absent. Moreover, in this case, the boundary
conditions (\ref{ADM}) do not play any role.

When $\rho <0$, the evaporation takes place but leaves a remnant. In fact,
Eq.(\ref{FM}) with $\overline{M}=0$ reads:
\begin{equation*}
\frac{1}{2}\partial _{r}\int_{0}^{T}\exp \left[ j(\tau ,r)\right] d\tau =%
\frac{M}{r-M}
\end{equation*}
which gives
\begin{equation}
\frac{1}{2}\int_{0}^{T}\exp \left[ j(\tau ,r)\right] d\tau =M\ln \left|
\frac{r}{M}-1\right| +N,  \label{zero}
\end{equation}
where $N$ is an integration constant.

The above relation is not compatible with the boundary conditions because:
\begin{equation*}
\infty >\frac{1}{2}Te^{K_{1}}\geq \frac{1}{2}\int_{0}^{T}e^{j(\tau ,r)}d\tau
\underset{r\rightarrow +\infty }{\rightarrow }\infty \text{,}
\end{equation*}
so that, even with the ''worse'' energy-momentum tensor (\textit{i.e.} a $%
T_{\mu \nu }$ violating all the energy conditions), $\overline{M}$ cannot
vanish, that is, the evaporation stops due to the trace anomaly of the
scalar fields that carry away energy from the blackhole to infinity.
Furthermore, by requiring a finite \textit{total Bondi flux }\cite{Bo62}%
\textit{,} it is easy to show that the same results also hold in the limit $%
T\rightarrow \infty $.

In conclusion, the \textit{information} \textit{loss} \textit{paradox} could
be overcome by taking into account the informations detained by the remnant
and by the Hawking radiation through the Bekenstein mechanism \cite{Be93}.
This is very appealing since the key ingredient to stop the evaporation,
namely the trace anomaly, is a typical feature of QFT, so QFT itself could
care for the loss of unitarity and this would be a good self-consistency
test for QFT.

This result looks quite surprising. In fact, many works on the back reaction
\cite{Ba81},\cite{Ma95} suggest that the blackhole evaporates adiabatically
in a time $O(M^{3})$ as first predicted by Hawking \cite{Ha74}. However, the
four dimensional trace anomaly brings in an important new feature which give
rise to the question: \textit{when is the trace anomaly not anymore
negligible?} This question, as we will now explain, cannot be answered by
only looking at Eqs. (\ref{F1}) and (\ref{F2}).

\textit{Trace anomaly, cosmological constant and scale invariance \ }Let us
observe that Eqs (\ref{F1}) and (\ref{F2}) are formally scale-invariant.
Namely, if $m_{0}$ and $j_{0}$ are a solution of the system, then $\lambda
m_{0}$ and $j_{0}$ are a solution too (provided we also make the
transformation $r\rightarrow \lambda r$, $t\rightarrow \lambda t)$. Thus,
Eqs (\ref{F1}) and (\ref{F2}) by themselves do not provide us with a lower
bound on $\overline{M}$.

However, the scale invariance is broken by the introduction of a
nonvanishing cosmological constant since scale transformations on $\Lambda $
cannot be performed, being $\Lambda $ a physical measurable parameter whose
value is fixed by observations. Roughly speaking, this means that $\Lambda $
introduces a characteristic energy scale. The same is true for the trace
anomaly, because it also breaks the scale invariance.

To get a lower bound on $\overline{M}$ and a new insight into the mechanism
stopping the evaporations, we can proceed as follows.

\textit{Thermodynamical features \ \ }As originally found by Hawking \cite
{Ha74}, the expectation value of the operator number, as measured by an
asymptotically static observer in a Schwarzschild blackhole of mass $M$, is
\begin{align}
n_{i}& =\frac{1}{\exp (\hbar \omega _{i}\sigma M)-1}  \label{radhaw} \\
\sigma & =\frac{\kappa }{\hbar ^{2}c^{3}}  \notag
\end{align}
where, in this formulas, the constants $\hbar $ and $c$ have been restored.
Since only a magnitude order estimate will be performed, in the following
the \textit{greybody} \textit{factors} will be neglected.

Relation (\ref{radhaw}) basically follows from the existence of a horizon
and from the fact that the notion of a particle is not an invariant concept
on curved space-time. Thus, in some sense, the Hawking effect is only
''kinematical'' \cite{Vi98}, since the Einstein equations are not needed to
get the relation (\ref{radhaw}). This obviously means that, in deducing
relation (\ref{radhaw}), we are neglecting the backreaction effects of the
emitted particles on the metric.

To take into account the backreaction, we have to impose the conservation of
energy, \textit{i.e.} that this thermal particles are created by the mass
lost by the blackhole. In the following, it will be assumed that the
expectation value of the operator number $n_{i}=n_{i}\left( M\right) $
varies adiabatically as a function of $M$, that is, if $M\rightarrow M\prime
$ then $n_{i}\rightarrow n_{i}\prime =n_{i}\left( M\prime \right) $. This
assumption is well verified in all situations in which the blackhole
thermodynamics is applicable \cite{Ma95}. Therefore, if $M\rightarrow
M+\delta M$, then $n_{i}\rightarrow n_{i}+\delta n_{i}$ with
\begin{equation*}
\delta n_{i}=\frac{\partial n_{i}}{\partial M}\delta M=-\frac{\hbar \omega
_{i}\exp (\hbar \omega _{i}\sigma M)}{\left( \exp (\hbar \omega _{i}\sigma
M)-1\right) ^{2}}\sigma \delta M\text{.}
\end{equation*}
To assure that the new particles are created by the mass lost by the
black-hole, we have to impose the conservation of energy:
\begin{equation}
c^{2}\delta M+\sum_{i}\hbar \omega _{i}\delta n_{i}\leq 0,  \label{rad1}
\end{equation}
where, since the blackhole is evaporating, $\delta M<0,$ and in Eq. (\ref
{rad1}) the inequality has been used to take into account the fact that the
mass lost by the blackhole gives rise to other kind of particles (such as
fermions, bosons, etc.) too. Then, the second term in Eq. (\ref{rad1})
encodes, in the thermodynamical limit, the backreaction effects of the
emitted particles on the gravitational field. It is clear from Eq. (\ref
{rad1}) that, as long as $M$ is big enough, the second term in the
inequality is completely negligible and the usual blackhole thermodynamics
should apply.

From Eq. (\ref{rad1}) we obtain
\begin{equation}
c^{2}\geq \sigma \underset{i}{\sum }\frac{\left( \hbar \omega _{i}\right)
^{2}\exp (\hbar \omega _{i}\sigma M)}{\left( \exp (\hbar \omega _{i}\sigma
M)-1\right) ^{2}}\text{.}  \label{rad2}
\end{equation}
The above relation, in which neither infrared nor ultraviolet divergences
appear, implies that the conservation of energy can be satisfied only for $%
M\geq \overline{M}$, where $\overline{M}$ is implicitly defined by the
following equation:
\begin{equation}
c^{2}=\sigma \underset{i}{\sum }\frac{\left( \hbar \omega _{i}\right)
^{2}\exp (\hbar \omega _{i}\sigma \overline{M})}{\left( \exp (\hbar \omega
_{i}\sigma \overline{M})-1\right) ^{2}}\text{.}  \label{rad3}
\end{equation}
Of course, $\overline{M}$\ is of the order of the Planck mass.

Thus, the conservation of energy, and then the backreaction, prevent the
blackhole mass $M$ from being lower then $\overline{M}$, so the remnant is
stable, since a smaller mass is not allowed by the energy conservation. This
conclusion clarifies the relation between energy conservation and complete
evaporation. In fact, if one tries to generalize QFT to allow nonunitary
processes, such as a complete evaporation, then the energy conservation is
lost \cite{El84}, \cite{Ba84}. Moreover, this, besides to be very
interesting in itself, fits very well with the numerical results obtained by
R. Casadio \cite{Ca01} in a Hamiltonian framework.

\section*{Acknowledgments}

The authors wish to thank D.Grumiller, G.Esposito and S.Sonego for remarks
and bibliographic suggestions.

\end{document}